\documentclass[10pt,a4paper]{article}
\usepackage[utf8x]{inputenc}
\usepackage{ucs}
\usepackage[english]{babel}
\usepackage{amsmath}
\usepackage{amsfonts}
\usepackage{amssymb}
\usepackage[left=2.5cm,right=2.5cm,top=2.5cm,bottom=2.5cm]{geometry}
\usepackage{bbm}
\usepackage{times}
\usepackage{amsthm}
\usepackage{listings,xcolor}

\usepackage{authblk}

\title{Joint likelihood calculation for intervention and observational data from a Gaussian Bayesian network}

\author[1]{G. Nuel}
\author[2]{A. Rau}
\author[2]{F. Jaffr{\'e}zic}
{ \small
\affil[1]{MAP5, UMR CNRS 8145, University Paris Descartes, Paris, France}
\affil[2]{INRA, UMR 1313 GABI, Jouy-en-Josas, France}
}

\begin{document}

\maketitle

\begin{abstract}
Methodological development for the inference of gene regulatory networks from transcriptomic data is an active and important research area. Several approaches have been proposed to infer relationships among genes from observational steady-state expression data alone, mainly based on the use of graphical Gaussian models. However, these methods rely on the estimation of partial correlations and are only able to provide undirected graphs that cannot highlight causal relationships among genes. A major upcoming challenge is to jointly analyze observational transcriptomic data and intervention data obtained by performing knock-out or knock-down experiments in order to uncover causal gene regulatory relationships. To this end, in this technical note we present an explicit formula for the likelihood function for any complex intervention design in the context of Gaussian Bayesian networks, as well as its analytical maximization. This allows a direct calculation of the causal effects for known graph structure. We also show how to obtain the Fisher information in this context, which will be extremely useful for the choice of optimal intervention designs in the future.\\
\\
\noindent {\bf Keywords}: Gaussian Bayesian networks, causal effects, intervention data, Fisher information.
\end{abstract}

\section{Introduction}

Inference of gene regulatory networks from transcriptomic data has been a wide research area in recent years. Several approaches have been proposed, mainly based on the use of graphical Gaussian models \cite{Friedman2008}. These methods, however, rely on the estimation of partial correlations and provide undirected graphs that cannot highlight the causal relationships among genes. B{\"u}hlmann {\it et al.} \cite{Maathuis2010,Buhlmann2009} recently proposed a method to predict causal effects from observational data alone in the context of Gaussian Bayesian networks (GBN). In this method, the PC algorithm \cite{Kalisch2012} is first applied to find the associated complete partially directed acyclic graph (CPDAG) among the graphs belonging to the corresponding equivalence class. Then, intervention calculus \cite{Pearl2000} is performed to estimate bounds for total causal effects based on each directed acyclic graph (DAG) in the equivalence class.

If knock-out or knock-down experiments are available, however, it is valuable to perform causal network inference from a mixture of observational and intervention data. One approach has been proposed to do so \cite{Pinna2010}, based on a simple comparison of observed gene expression values to the expression under intervention; the underlying idea is that if gene $Y$ is regulated by gene $X$, then its expression value under a knock-out of gene $X$ will be different from the value in a wild type experiment. We note that this method provided the best network estimation in the DREAM4 challenge, and has the advantage of being very fast to compute without imposing a restriction on the acyclicity of the graph. It does, however, require an intervention experiment to be performed for each gene, which can be unrealistic for real applications given the cost and time typically involved for knock-out experiments.  In addition, although it is well-suited to the inference of the structure of the graph, it tends to be imprecise for the estimation of the strength of the interactions between genes.

The aim of this technical note is to propose an explicit calculation of the likelihood function for complex intervention designs, including both observational and intervention data, in the context of GBNs. This calculation makes use of the full set of available information available, does not require an intervention for each gene, and is able to deal with multiple interventions (e.g., a double gene knock-out experiment). For an known graph structure, we present here the likelihood calculation for observational data only, as well as for any intervention design. We also provide the analytical first order derivatives which allow a direct estimation of the graph structure as well as the causal effects. Finally, we give the Fisher information, which is not trivial to derive, and will be extremely useful in the future for the choice of optimal intervention designs.

The rest of this technical note is organized as follows. In Section 2, we define the model and set up a toy example for illustrative purposes. In Sections 3 and 4, we define the likelihood function, maximum likelihood estimators, and Fisher information in the case where only observational data are available and in the case where a mixture of intervention and observational data are available, respectively. In Section 5, we provide a brief discussion and conclusion.

\section{Model definition}

\subsection{Definition} 
We consider the set $X_\mathcal{I}$, $\mathcal{I}=\{1,\ldots,p\}$ a set of $p$ Gaussian random variables defined by:
\begin{align}
X_j=m_j+\sum_{i \in \text{pa}(j)} w_{i,j} X_i + \varepsilon_j
\quad
\text{with}
\quad \varepsilon_j \sim \mathcal{N}(0,\sigma_j^2). \label{eqn:model}
\end{align}
We assume that the $\varepsilon_j$ are independent, and that $i \in \text{pa}(j) \Rightarrow i<j$; this assumption is equivalent to assuming that the directed graph obtained using the parental relationships is acyclic. Given the parental structure of the graph, the model parameters are $\theta=(m,\sigma,w)$ where $w_{i,j}$ is nonzero only on the edge set $(i,j)\in \mathcal{E}=\{i \in \text{pa}(j),j \in \mathcal{I} \}$.

It is easy to see that this model is equivalent to $X_\mathcal{I} \sim \mathcal{N}( \boldsymbol{\mu} ; \boldsymbol{\Sigma})$, with:
$$
\boldsymbol{\mu}=m  \mathbf{L}
\quad\text{and}\quad
\boldsymbol{\Sigma}= \mathbf{L}^T\text{diag}(\sigma^2)\mathbf{L}= \sum_{j} \sigma_j^2 \mathbf{L}^T e_j^T e_j \mathbf{L}
$$
where $e_j$ is a null row-vector except for the its $j^\text{th}$ term which is equal to $1$, and where  $\mathbf{L}=(\mathbf{I}-\mathbf{W})^{-1}=\mathbf{I}+\mathbf{W}+\ldots+\mathbf{W}^{p-1}$ with $\mathbf{W}=(w_{i,j})_{i,j \in \mathcal{I}}$. Note that the nilpotence of $\mathbf{W}$ is due to the fact that $w_{i,j}=0$ for all $i\geqslant j$.

\subsection{A toy example}

We consider the particular case where $p=3$, $\text{pa}(1)=\emptyset$, $\text{pa}(2)=\{1\}$,  $\text{pa}(3)=\{1,2\}$; the true values of the parameters are set to be $m^*=(0.5\ 1.2\ 0.7)$; $\sigma^*=(0.3\ 1.1\ 0.6)$; $w_{1,2}^* =-0.8$, $w_{1,3}^*=0.9$, and $w_{2,3}^*=0.5$.  We hence have
$$
\mathbf{W}=\left(
\begin{array}{ccc}
0 & -0.8 & 0.9 \\
0 & 0 & 0.5 \\
0 & 0 & 0
\end{array}
\right)
\quad\text{and}\quad
\mathbf{L}=\mathbf{I}+\mathbf{W}+\mathbf{W}^2=\left(
\begin{array}{ccc}

1.0 & -0.8 & 0.5 \\
0 & 1.0 & 0.5 \\
0 & 0 & 1.0
\end{array}
\right)
$$
and observed data can be generated through $X_{1:3}\sim \mathcal{N}(\boldsymbol{\mu};\boldsymbol{\Sigma})$ with:
$$
\boldsymbol{\mu}=(0.5\ 0.8\ 1.55)
\quad\text{and}\quad
\boldsymbol{\Sigma}=\left(
\begin{array}{ccc}
  0.090 &-0.0720 &0.045 \\
 -0.072 & 1.2676 &0.569 \\
 0.045  &0.5690 &0.685 \\
\end{array}
\right).
$$

\section{Observational data}

\subsection{Likelihood}

The log-likelihood of the model described in Equation~(\ref{eqn:model}), given $N$ observations $x^k=(x^k_1,\ldots,x^k_p)$ ($1 \leqslant k \leqslant N$), may be written as follows:
\begin{equation}
\ell(m,\sigma,w)=-\frac{Np}{2} \log (2\pi)-N\sum_{j} \log (\sigma_j) 
-\frac{1}{2} \sum_{j} \frac{1}{\sigma_j^2} \sum_{k} (x^k_j - x^k\mathbf{W} e_j^T -m_j)^2.
\label{eq:likelihood_obs}
\end{equation}
\begin{proof}
For all $k$, let us define $A_k= (x^k-m\mathbf{L}) \boldsymbol{\Sigma}^{-1} (x^k-m\mathbf{L})^T$. Since
$\boldsymbol{\Sigma}^{-1}=(\mathbf{I}-\mathbf{W})\text{diag}(1/\sigma^2)(\mathbf{I}-\mathbf{W})^T$ we obtain:
\begin{eqnarray*}
A_k &=& \sum_{j} \frac{1}{\sigma_j^2} (x^k(\mathbf{I}-\mathbf{W})-m)e_j^T e_j (x^k(\mathbf{I}-\mathbf{W})-m)^T\\
&=&\sum_{j} \frac{1}{\sigma_j^2} (x^k_j - x^k\mathbf{W} e_j^T -m_j)^2.
\end{eqnarray*}
\end{proof}

We now consider the derivative of $\ell$ with respect to $m$:
$$
\frac{\partial \ell}{\partial m_j}(m,\sigma,w)
= \frac{1}{\sigma_j^2} \sum_{k}  (x^k_j -x^k\mathbf{W}e_j^T -m_j).
$$
The maximization of $\ell(m,\sigma,w)$ in $m$ for a fixed $w$ hence does not depend on $\sigma$ and is given by:
\begin{equation*}
\hat{m}_j=\frac{1}{N} \sum_{k=1}^N (x^k_j -x^k\mathbf{W}e_j^T).
\end{equation*}
By replacing $m_j$ with this formula in Equation~(\ref{eq:likelihood_obs}) we get an expression of the likelihood free of the parameter $m$:
\begin{align}
\tilde{\ell}(\sigma,w)=-\frac{Np}{2} \log (2\pi)-N\sum_{j} \log (\sigma_j) 
-\frac{1}{2} \sum_{j} \frac{1}{\sigma_j^2}\sum_{k} (y^k_j - y^k\mathbf{W} e_j^T)^2 \label{eq:likelihood_obs2}
\end{align}
where for all $k,j$ we have:
\begin{equation}
y^k_j=x^k_j-\frac{1}{N}\sum_{k'} x^{k'}_j.
\label{eq:yk}
\end{equation}

\subsection{Maximum likelihood estimator}

\subsubsection{Derivatives with respect to $w$}

The derivatives of $\tilde{\ell}$ defined in Equation~(\ref{eq:likelihood_obs2}) with respect to $w$ are as follows:
$$
\frac{\partial \tilde{\ell}}{\partial w_{i,j}}(\sigma,w)=
\frac{1}{\sigma_{j}^2}\sum_{k} y^k_i (y^k_j- y^k\mathbf{W}e_j^T).
$$
\begin{proof}
We first note the following:
$$
\sum_{j'} \frac{1}{\sigma_{j'}^2} (y^k_{j'} - y^k\mathbf{W} e_{j'}^T)(\underbrace{y^ke_i^T}_{y^k_i}\underbrace{e_j e_{j'}^T}_{\mathbbm{1}_{j'=j}})
=\frac{y_j^k}{\sigma_{j}^2} (y^k_{j} - y^k\mathbf{W} e_{j}^T).
$$
As such, the maximization of $\tilde{\ell}(\sigma,w)$ in $w$ can be done independently from $\sigma$ by solving for all $(i,j)\in \mathcal{E}$:
$$
\sum_{k=1}^N y^k_i y^k\mathbf{W}e_j^T= \sum_{k=1}^N y^k_i y^k_j.
$$
Hence using
$$
\mathbf{W}=\sum_{(i',j')\in \mathcal{E}} w_{i',j'} e_{i'}^T e_{j'}
\Rightarrow
y^k\mathbf{W}e_j^T = \sum_{i', (i',j)\in \mathcal{E}} w_{i',j} y^k_{i'} 
$$
we find that $\hat w$ is solution of the following linear system: 
\begin{equation*}
\sum_{i', (i',j)\in \mathcal{E}} \hat w_{i',j}\sum_{k=1}^N y_i^ky_{i'}^k = \sum_{k=1}^N y_i^k y_j^k
\quad\text{for all $(i,j)\in \mathcal{E}$}.
\end{equation*}
\end{proof}

\subsubsection{Derivatives with respect to $\sigma$}

The derivatives of $\tilde{\ell}$  defined in Equation~(\ref{eq:likelihood_obs2}) with respect to $\sigma$ are:
$$
\frac{\partial \tilde{\ell}}{\partial \sigma_j}(\sigma,w)=-\frac{N}{\sigma_j}
+\frac{1}{\sigma_j^3}\sum_{k} (y^k_j - y^k\mathbf{W}e_j^T)^2.
$$
The maximization of $\ell(\sigma,w)$ in $\sigma$ when $m$ is fixed is thus given by:
\begin{equation*}
\hat\sigma_j^2 
= \frac{1}{N} \sum_{k} (y^k_j - y^k\mathbf{W}e_j^T)^2.
\end{equation*}

\subsubsection{Maximum of likelihood}

If we now plug the MLE expressions back into the likelihood we get:
$$
\max_{m,\sigma,w} \ell(m,\sigma,w)=
\frac{Np}{2} (\log N - \log(2\pi) - 1)
+ \max_{w} \left\{ -\frac{N}{2}\sum_j \log \left( \sum_k (y_j^k-y^k \mathbf{W} e_j^T)^2 \right) \right\}
$$
which means that the only part of this expression that is connected to the observed values is: $S_j=\sum_k (y_j^k-y^k \mathbf{W} e_j^T)^2$ for all $j$. If we now assume that the model is full, which means $w_{i,j} \neq 0$ for all $i<j$, we get
$$
\sum_k (y_j^k-y^k \mathbf{W} e_j^T)^2
= Y_{j,j} - 2 \sum_{i<j} w_{i,j} Y_{i,j} + \sum_{i<j} \sum_{i'<j} w_{i,j}w_{i',j} Y_{i,i'}
$$
where $Y_{i,j}=\sum_k y_i^k y_j^k$ for all $i,j$. Taking advantage of the relation: $\sum_{i'<j} w_{i',j} Y_{i,i'}=Y_{i,j}$ we finally get
$$
S_j
=Y_{j,j} -  \sum_{i<j} w_{i,j} Y_{i,j}
=Y_{j,j} - \mathbf{b_{j-1}^T} \mathbf{A}_{j-1}^{-1} \mathbf{b_{j-1}}
$$
where $\mathbf{b}_{j-1}=(Y_{i,j})_{i<j}$ and $\mathbf{A}_{j-1}=(Y_{i,i'})_{i,i'<j}$.

It is now easy to prove by recurrence that:
$$
\prod_{j=1}^p S_j = \det\left(\mathbf{A}_{p+1}\right)
$$
the key point being the following relationship (this is a result of basic linear algebra using the product of block-trigonal matrices):
$$
\det\left(
\mathbf{A}_{p+1}
\right)=
\det\left(
\begin{array}{cc}
\mathbf{A}_p & \mathbf{b}_p \\
\mathbf{b}_p^T & Y_{p,p}
\end{array}
\right)=\det(\mathbf{A}_p)\det( Y_{p,p} - \mathbf{b}_p^T\mathbf{A}_{p}^{-1} \mathbf{b}_p).
$$

Thanks to this result, in the particular case of the full model ($w_{i,j}\neq 0$ for all $i<j$) we hence have:
$$
\max_{m,\sigma,w} \ell(m,\sigma,w)=
\frac{Np}{2} (\log N - \log(2\pi) - 1)
 -\frac{N}{2} \log \det \left( \mathbf{A}_{p+1} \right)
$$
and since the determinant is invariant to any permutation of the row and columns, this maximum is the same for all possible orderings of the variables $1,\ldots,p$.

\subsubsection{Toy example (continued)}

Here is a sample of size $N=5$ drawn from our toy example model:
$$
x=
\left(
\begin{array}{ccc}
1.1025540 &-0.2652622 &1.957083\\
0.6721755 & 0.4286717 &1.605024\\
0.3455340 & 2.8835932 &1.932982\\
0.4139627 & 1.0847936 &1.250889\\
0.2844364 & 1.0490652 &1.446954\\
\end{array}
\right)
$$
where after centering we get:
$$
y=
\left(
\begin{array}{ccc}
0.5388215 &-1.30143445 & 0.31849676\\
0.1084430 &-0.60750058 &-0.03356279\\
-0.2181985 & 1.84742088 & 0.29439582\\
-0.1497698 & 0.04862127 &-0.38769719\\
-0.2792962 & 0.01289289 &-0.19163260\\
\end{array}
\right).
$$
We obtain $\hat w$ by solving:
$$
\left(
\begin{array}{ccc}
0.4501364 & 0.0000000 & 0.000000\\
0.0000000 & 0.4501364 &-1.181107\\
0.0000000 &-1.1811075 & 5.478283\\
\end{array}
\right)
\left(
\begin{array}{c}
\hat w_{1,2} \\
\hat w_{1,3} \\
\hat w_{2,3} \\
\end{array}
\right)=
\left(
\begin{array}{c}
-1.1811075 \\ 0.2153241 \\ 0.1284387
\end{array}
\right)
$$
which gives (reference value in parentheses):
$$
\hat w_{1,2} = -2.6238878\ (w_{1,2}^*=-0.8)
\quad
\hat w_{1,3} = 1.2430964\ (w_{1,3}^*=0.9)
\quad
\hat w_{2,3} = 0.2914543\ (w_{2,3}^*=0.5).
$$
Finally, we then obtain:
$$
\hat \sigma_1= 0.3000455\ (\sigma_{1}^*=0.3)
\quad
\hat \sigma_2= 0.6898100\ (\sigma_{2}^*=1.1)
\quad
\hat \sigma_3= 0.1193022\ (\sigma_{3}^*=0.6)
$$
and the nuisance parameter:
$$
\hat m_1=0.5637325\ (m_{1}^*=0.5)
\quad
\hat m_2=2.5153432\ (m_{2}^*=0.2)
\quad
\hat m_3=0.6358156\ (m_{3}^*=0.7).
$$

Finally we check that:
$$
\log\det \left(
\mathbf{A}_{4}
\right)
=
\log\det \left(
\begin{array}{ccc}
 0.450136 & -1.181107 & 0.215324\\
-1.181107 &  5.478283 & 0.128439\\
 0.215324 &  0.128439 & 0.376268\\
\end{array}
\right)
=-2.574198
=2 \sum_j \log(\hat\sigma_j)+p\log N.
$$

\subsection{Fisher information}

\subsubsection{Hessian of $\tilde{\ell}$}

The (non-zero) second order derivatives of $\tilde{\ell}$ are given by:
$$
\frac{\partial^2 \tilde{\ell}}{\partial w_{i,j}\partial w_{i',j}}(\sigma,w)=-
\frac{1}{\sigma_{j}^2}\sum_{k} y^k_i y^k_{i'}
\quad
\frac{\partial^2 \tilde{\ell}}{\partial w_{i,j}\partial \sigma_{j}}(\sigma,w)=
-\frac{2}{\sigma_{j}^3}\sum_{k} y^k_i (y^k_j- y^k\mathbf{W}e_j^T).
$$
$$
\frac{\partial^2 \tilde{\ell}}{\partial \sigma_j^2}(\sigma,w)=\frac{N}{\sigma_j^2}
-\frac{3}{\sigma_j^4}\sum_{k} (y^k_j - y^k\mathbf{W}e_j^T)^2.
$$

\subsubsection{Distribution of $y^ k$}

We can rewrite Equation~(\ref{eq:yk}) as:
$$
y^k=\frac{N-1}{N} x^k - \frac{1}{N}\sum_{k' \neq k} x^{k'}
$$
which is obviously a Gaussian vector with expectation:
$$
\mathbb{E}[y^k]=\frac{N-1}{N} \boldsymbol{\mu} - \sum_{k' \neq k}\boldsymbol{\mu} = \left(\frac{N-1}{N} - \frac{N-1}{N}\right)\boldsymbol{\mu}=  {\bf 0}
$$
and variance:
$$
\mathbb{V}[y^k]=\frac{(N-1)^2}{N^2} \boldsymbol{\Sigma}
+ \frac{N-1}{N^2} \boldsymbol{\Sigma} = \frac{N-1}{N} \boldsymbol{\Sigma}.
$$
It is therefore easy to establish that:
\begin{equation}
y^k \sim \mathcal{N}\left(
\mathbf{0};
\frac{N-1}{N}  \boldsymbol{\Sigma}
\right)
\quad\text{and}\quad
y^k - y^k \mathbf{W} \sim \mathcal{N}\left(
\mathbf{0};
\frac{N-1}{N} \text{diag}(\sigma^2)
\right).
\end{equation}
Note that as a consequence of this, it is easy to prove that $\mathbb{E}[\hat \sigma_j^2]=(N-1)/N \sigma_j^2$ which means that this estimator is (slightly) biased.

\subsubsection{Information}

The Fisher information matrix $\mathbf{I}(\sigma,w)$ can therefore be written as:
$$
e_{w_{i,j}}\mathbf{I}(\sigma,w) e_{w_{i',j}}^T=-\mathbb{E}\left[
\frac{\partial^2 \tilde{\ell}}{\partial w_{i,j}\partial w_{i',j}}(\sigma,w)
\right]
=\frac{1}{\sigma_j^2} \sum_k \mathbb{E}[y_i^k y_{i'}^k]
=\frac{N-1}{\sigma_j^2} \boldsymbol{\Sigma}_{i,i'}
$$
$$
e_{w_{i,j}}\mathbf{I}(\sigma,w) e_{\sigma_j}^T=-\mathbb{E}\left[
\frac{\partial^2 \tilde{\ell}}{\partial w_{i,j}\partial \sigma_j}(\sigma,w)
\right]
=\frac{2}{\sigma_j^3} \sum_k e_i \mathbb{E}\left[(y^k)^T y^k\right] (\mathbf{I}-\mathbf{W}) e_j^T
=\frac{2(N-1)}{\sigma_j^3} e_i \boldsymbol{\Sigma}(\mathbf{I}-\mathbf{W}) e_j^T
=0
$$
because $e_i \boldsymbol{\Sigma}(\mathbf{I}-\mathbf{W}) e_j^T=\sigma_j^2e_i \mathbf{L}^T e_j^T=\sigma_j^2 e_j \mathbf{L} e_i^T$ and $\mathbf{L}$ is upper triangular. And finally:
$$
e_{\sigma_j}\mathbf{I}(\sigma,w) e_{\sigma_j}^T=
-\mathbb{E}\left[
\frac{\partial^2 \tilde{\ell}}{\partial w_{i,j}\partial \sigma_j}(\sigma,w)
\right]=
\frac{3}{\sigma_j^4} \sum_k \mathbb{E}\left[(y_j^k-y^k\mathbf{W}e_j^T)^2 \right]-\frac{N}{\sigma_j^2}
=\frac{3(N-1)}{\sigma_j^2}-\frac{N}{\sigma_j^2}
=\frac{2N-3}{\sigma_j^2}.
$$

\subsubsection{Toy example (continued)}

We present here the inverse Fisher information matrix in the particular case of our toy example model. Due to the block-wise nature of $\mathbf{I}(\sigma,w)$, the Cramer-Rao lower bound on the covariance matrix is given by blocks:
$$
(N-1)\mathbb{V}\text{ar}(\hat w_{1,2})= \frac{\sigma_2^2}{\boldsymbol{\Sigma}_{1,1}}=13.444
$$
$$
(N-1)\mathbb{C}\text{ov}(\hat w_{1,3},\hat w_{2,3})=
\sigma_3^2
\left(
\begin{array}{cc}
\boldsymbol{\Sigma}_{1,1} & \boldsymbol{\Sigma}_{1,2} \\
\boldsymbol{\Sigma}_{2,1} & \boldsymbol{\Sigma}_{2,2} \\
\end{array}
\right)^{-1}
=
\left(
\begin{array}{cc}
4.1904132 &0.2380165\\
0.2380165 &0.2975207\\
\end{array}
\right)
$$
$$
(2N-3)\mathbb{C}\text{ov}(\hat \sigma_1,\hat \sigma_2,\hat \sigma_3)=
\text{diag}(\sigma^2)=
\text{diag}(0.09,1.21,0.36).
$$

In the particular case where $N=200$, the standard deviations corresponding to the Cramer-Rao bounds for $\hat\theta=(\hat w_{1,2},\hat w_{1,3},\hat w_{2,3},\hat \sigma_1,\hat \sigma_2,\hat \sigma_3)$ are:
$$
\text{sd}_\text{CR}(\hat\theta)=
\left(
\begin{array}{cccccc}
0.25992311 &0.14511152 &0.03866625 &0.01505657 &0.05520742 &0.03011314
\end{array}
\right)
$$
while the empirical standard-deviation (sample size 2000) are:
$$
\text{sd}_\text{emp}(\hat\theta)=
\left(
\begin{array}{cccccc}
0.26110572 &0.14847838 &0.03873625 &0.01476351 &0.05424152 &0.02952547
\end{array}
\right).
$$
The empirical mean is:
$$
\text{mean}_\text{emp}(\hat\theta)=
\left(
\begin{array}{cccccc}
-0.8024838 & 0.8996667 & 0.5004233 & 0.2989493 & 1.0935090 & 0.5941743 
\end{array}
\right)
$$
while the true parameter is:
$$
\theta^*=
\left(
\begin{array}{cccccc}
-0.80 & 0.90 & 0.50 & 0.30 & 1.10 & 0.60
\end{array}
\right).
$$

\section{Mixture of intervational and observational data}

\subsection{Case of a single intervention experiment}

We assume now that we perform an intervention on a subset $\mathcal{J} \subset \mathcal{I}=\{1,\ldots,p\}$ of variables by artificially setting the level of the corresponding variables to a value: $\text{do}(X_\mathcal{J}=x_\mathcal{J})$. The corresponding model is obtained by assuming that all $w_{i,j}=0$ for $(i,j) \in \mathcal{E}$ and $j \in \mathcal{J}$; we denote the corresponding matrix $\mathbf{W}_\mathcal{J}$. We also assume that the variables $X_j$ for $j \in \mathcal{J}$ are fully deterministic. The resulting model is hence Gaussian once again: $X_{\mathcal{I}} | \text{do}(X_\mathcal{J}=x_\mathcal{J}) \sim \mathcal{N}(\boldsymbol{\mu}_{\mathcal{J}}(x_\mathcal{J}),
\boldsymbol{\Sigma}_{\mathcal{J}})$ with
$$
\boldsymbol{\mu}_{\mathcal{J}}(x_\mathcal{J})=
\boldsymbol{\nu}_\mathcal{J}(x_\mathcal{J}) \mathbf{L}_{\mathcal{J}}
,\quad
\boldsymbol{\Sigma}_{\mathcal{J}}=   \mathbf{L}_{\mathcal{J}}^T \text{diag}(\sigma^2) D_\mathcal{J} \mathbf{L}_{\mathcal{J}} 
$$
where $D_\mathcal{J}=\sum_{j \notin \mathcal{J}} e_j^Te_j$ is a diagonal matrix with $0$ at $\mathcal{J}$ positions and $1$ elsewere, and with
$$
\boldsymbol{\nu}_\mathcal{J}(x_\mathcal{J})e_j^T=\left\{
\begin{array}{ll}
x_j & \text{if $j \in \mathcal{J}$} \\
m_j & \text{else} \\
\end{array}
\right.
\quad\text{and}\quad
\mathbf{L}_{\mathcal{J}}=(\mathbf{I}-\mathbf{W}_{\mathcal{J}})^{-1}=\mathbf{I}+\mathbf{W}_{\mathcal{J}}+\ldots+\mathbf{W}_{\mathcal{J}}^{p-1}.
$$

\subsection{Maximum likelihood estimator}

\subsubsection{Likelihood}

We consider $N$ data generated under $x^k=(x^k_1,\ldots,x^k_p)$ ($1 \leqslant k \leqslant N$) with intervention on $\mathcal{J}_k$ ($\mathcal{J}_k=\emptyset$ means no intervention). We denote by $\mathcal{K}_j=\{k,j \notin \mathcal{J}_k\}$, and by $N_j=|\mathcal{K}_j|$ its cardinal. The log-likelihood of the model can then be written as:
\begin{equation*}
\ell(m,\sigma,w)=-\frac{\log(2\pi)}{2}\sum_j N_j -\sum_{j} N_j \log (\sigma_j)
-\frac{1}{2} \sum_j \frac{1}{\sigma_j^2}\sum_{k \in \mathcal{K}_j}  (x^k_j-x^k\mathbf{W}e_j^T-m_j)^2.
\end{equation*}
\begin{proof}
This is mainly due to the fact that for any intervention set $\mathcal{J}$ we have $\mathbf{W}_{\mathcal{J}}e_j^T=\mathbf{W}e_j^T$ for all $j \notin \mathcal{J}$.
\end{proof}

Considering the derivative with respect to $m_j$ we get for all $j$ such that $N_j>0$:
\begin{equation*}
\hat{m}_j=\frac{1}{N_j} \sum_{k \in \mathcal{K}_j} (x^k_j-x^k\mathbf{W}e_j^T)
\end{equation*}
which can be plugged into the likelihood expression to get:
\begin{equation*}
\tilde \ell(\sigma,w)=-\frac{\log(2\pi)}{2}\sum_j N_j-\sum_{j} N_j \log (\sigma_j)
-\frac{1}{2} \sum_{j} \frac{1}{\sigma_j^2} \sum_{k \in \mathcal{K}_j}  (y^{k,j}_j-y^{k,j}\mathbf{W}e_j^T)^2
\end{equation*}
where for $(k,j)$ such as $k \in \mathcal{K}_j$ we have:
$$
y^{k,j}=x^k-\frac{1}{N_j} \sum_{k'\in \mathcal{K}_j} x^{k'}.
$$

\subsubsection{Estimators}

It can be shown that $w$ may be estimated by solving the following linear system:
\begin{equation*}
\sum_{i', (i',j)\in \mathcal{E}} w_{i',j}\sum_{k \in \mathcal{K}_j} y_i^{k,j}y_{i'}^{k,j} = \sum_{k \in \mathcal{K}_j} y_i^{k,j} y_j^{k,j}
\quad\text{for all $(i,j)\in \mathcal{E}$}.
\end{equation*}
Note that the system might be degenerate if the intervention design gives no insight on some parameters.

It is hence finally possible to obtain an estimator of $\sigma$ through:
\begin{equation*}
\hat{\sigma}_j^2 
= \frac{1}{N_j} \sum_{k \in \mathcal{K}_j} (y^{k,j}_j - y^{k,j}\hat{\mathbf{W}}e_j^T)^2.
\end{equation*}

\subsubsection{Toy example (continued)}

Let us consider the following design: $\mathcal{J}_{1}=\{1\}$ with $x_1^1=-0.5$, $\mathcal{J}_{2}=\{2\}$ with $x_2^2=0.5$, $\mathcal{J}_{3}=\{3\}$ with $x_3^3=0.1$, $\mathcal{J}_{4}=\{1,2\}$ with $x_1^4=-1.5$ and $x_2^4=2.5$, and  $\mathcal{J}_{5}=\emptyset$ (no intervention). 

Here is a sample of size $N=5$ drawn from our toy-example model:
$$
x=
\left(
\begin{array}{ccc}
-0.50000000 & 0.9391031 & 0.7665494\\
 0.47655556 & 0.5000000 & 1.4537910\\
 0.09892252 & 1.2963643 & 0.1000000\\
-1.50000000 & 2.5000000 & 0.3326028\\
 0.36614988 & 1.1787898 & 1.9014714\\
\end{array}
\right)
$$
after centering we get:
$$
y^{\cdot,1}=
\left(
\begin{array}{ccc}
-0.81387599 &-0.0526149 &-0.3852047\\
 0.16267957 &-0.4917180 & 0.3020369\\
-0.21495346 & 0.3046462 &-1.0517541\\
-1.81387599 & 1.5082820 &-0.8191513\\
 0.05227389 & 0.1870718 & 0.7497173\\
\end{array}
\right)
\quad
y^{\cdot,2}=
\left(
\begin{array}{ccc}
-0.4883575 &-0.19898261 &-0.1561242\\
 0.4881981 &-0.63808574 & 0.5311174\\
 0.1105651 & 0.15827852 &-0.8226736\\
-1.4883575 & 1.36191426 &-0.5900708\\
 0.3777924 & 0.04070409 & 0.9787978\\
\end{array}
\right)
$$
$$
y^{\cdot,3}=
\left(
\begin{array}{ccc}
-0.2106764 &-0.34037011 &-0.3470542\\
 0.7658792 &-0.77947324 & 0.3401873\\
 0.3882462 & 0.01689102 &-1.0136037\\
-1.2106764 & 1.22052676 &-0.7810009\\
 0.6554735 &-0.10068341 & 0.7878678\\
\end{array}
\right).
$$

We get $\hat w$ by solving:
$$
\left(
\begin{array}{ccc}
0.3934448 & 0.000000 & 0.000000\\
0.0000000 & 2.526338 &-2.068933\\
0.0000000 &-2.068933 & 2.223253\\
\end{array}
\right)
\left(
\begin{array}{c}
\hat w_{1,2} \\
\hat w_{1,3} \\
\hat w_{2,3} \\
\end{array}
\right)=
\left(
\begin{array}{c}
0.1300524 \\ 1.7956242 \\-1.1795977
\end{array}
\right)
$$
which gives (reference value in parentheses): 
$$
\hat w_{1,2} = 0.3305481\ (w_{1,2}^*=-0.8)
\quad
\hat w_{1,3} = 1.1612114\ (w_{1,3}^*=0.9)
\quad
\hat w_{2,3} = 0.5500366\ (w_{2,3}^*=0.5).
$$
We then finally obtain:    
$$
\hat \sigma_1= 0.15853727\ (\sigma_{1}^*=0.3)
\quad
\hat \sigma_2= 0.08815595\ (\sigma_{2}^*=1.1)
\quad
\hat \sigma_3= 0.08745639\ (\sigma_{3}^*=0.6)
$$
and the nuisance parameter: %
$$
\hat m_1=0.3138760\ (m_{1}^*=0.5)
\quad
\hat m_2=1.1419342\ (m_{2}^*=0.2)
\quad
\hat m_3=0.7458125\ (m_{3}^*=0.7).
$$

\subsection{Fisher information}

\subsubsection{Hessian of $\tilde{\ell}$}

The (non-zero) second order derivatives of $\tilde{\ell}$ are given by:
$$
\frac{\partial^2 \tilde{\ell}}{\partial w_{i,j}\partial w_{i',j}}(\sigma,w)=-f
\frac{1}{\sigma_{j}^2}\sum_{k\in \mathcal{K}_j} y^{k,j}_i y^{k,j}_{i'}
\quad
\frac{\partial^2 \tilde{\ell}}{\partial w_{i,j}\partial \sigma_{j}}(\sigma,w)=
-\frac{2}{\sigma_{j}^3}\sum_{k \in \mathcal{K}_j} y^{k,j}_i (y^{k,j}_j- y^{k,j}\mathbf{W}e_j^T).
$$
$$
\frac{\partial^2 \tilde{\ell}}{\partial \sigma_j^2}(\sigma,w)=\frac{N}{\sigma_j^2}
-\frac{3}{\sigma_j^4}\sum_{k \in \mathcal{K}_j} (y^{k,j}_j - y^{k,j}\mathbf{W}e_j^T)^2.
$$

\subsubsection{Distribution of $y^{k,j}$}

For all $k$, let us adopt the following notation: $\mathbf{W}_k=\mathbf{W}_{\mathcal{J}_k}$, $\mathbf{L}_k=\mathbf{L}_{\mathcal{J}_k}$, $\boldsymbol{\nu}_k=\boldsymbol{\nu}_{\mathcal{J}_k}$, 
$\boldsymbol{\mu}_k=\boldsymbol{\mu}_{\mathcal{J}_k}$, and
$\boldsymbol{\Sigma}_k=\boldsymbol{\Sigma}_{\mathcal{J}_k}$. We can then rewrite Equation~(\ref{eq:yk}) as:
$$
y^{k,j}= x^k - \frac{1}{N_j}\sum_{k' \in \mathcal{K}_j} x^{k'}
=\frac{N_j-1}{N_j} x^k - \frac{1}{N_j}\sum_{k' \in \mathcal{K}_j, k' \neq k} x^{k'}
$$
with $x^k \sim \mathcal{N}\left(
\boldsymbol{\mu}_k;
\boldsymbol{\Sigma}_k
\right)$, from which we derive that:
\begin{equation*}
y^{k,j} \sim \mathcal{N}\left(
\underbrace{\boldsymbol{\mu}_k-\frac{1}{N_j}\sum_{k' \in \mathcal{K}_j} \boldsymbol{\mu}_{k'}}_{\mathbf{m}^{k,j}};
\underbrace{\frac{(N_j-1)^2}{N_j^2}  \boldsymbol{\Sigma}_k
+\frac{1}{N_j^2} \sum_{k' \in \mathcal{K}_j, k' \neq k} \boldsymbol{\Sigma}_{k'}}_{\mathbf{S}^{k,j}}
\right)
\quad\text{and}\quad
y^{k,j}(\mathbf{I} - \mathbf{W})e_j^T \sim \mathcal{N}\left(
0;
\frac{N_j-1}{N_j} \sigma_j^2
\right).
\end{equation*}

\subsubsection{Information}

The Fisher information matrix $\mathbf{I}(\sigma,w)$ can therefore be written as:
\begin{multline*}
e_{w_{i,j}}\mathbf{I}(\sigma,w) e_{w_{i',j}}^T=-\mathbb{E}\left[
\frac{\partial^2 \tilde{\ell}}{\partial w_{i,j}\partial w_{i',j}}(\sigma,w)
\right]
=\frac{1}{\sigma_j^2} \sum_{k\in\mathcal{K}_j} \mathbb{E}[y_i^{k,j} y_{i'}^{k,j}]\\
=\frac{1}{\sigma_j^2} \sum_{k\in\mathcal{K}_j}
\left(\boldsymbol{S}^{k,j}_{i,i'}+\mathbf{m}_i^{k,j}\mathbf{m}_{i'}^{k,j}
\right)
=\frac{1}{\sigma_j^2} \left(\frac{N_j^2+ N_j -1}{N_j^2} \sum_{k\in\mathcal{K}_j} \boldsymbol{\Sigma}_k
+\sum_{k\in\mathcal{K}_j} \mathbf{m}_i^{k,j}\mathbf{m}_{i'}^{k,j}
\right)
\end{multline*}
$$
e_{w_{i,j}}\mathbf{I}(\sigma,w) e_{\sigma_j}^T=-\mathbb{E}\left[
\frac{\partial^2 \tilde{\ell}}{\partial w_{i,j}\partial \sigma_j}(\sigma,w)
\right]
=\frac{2}{\sigma_j^3} \sum_{k\in\mathcal{K}_j} \mathbb{E}\left[y_i^{k,j} y^{k,j} (\mathbf{I}-\mathbf{W}) e_j^T\right]
=0
$$
and finally:
$$
e_{\sigma_j}\mathbf{I}(\sigma,w) e_{\sigma_j}^T=
-\mathbb{E}\left[
\frac{\partial^2 \tilde{\ell}}{\partial w_{i,j}\partial \sigma_j}(\sigma,w)
\right]=
\frac{3}{\sigma_j^4} \sum_{k\in\mathcal{K}_j} \mathbb{E}\left[(y_j^{k,j}-y^{k,j}\mathbf{W}e_j^T)^2 \right]-\frac{N_j}{\sigma_j^2}
=\frac{3(N_j-1)}{\sigma_j^2}-\frac{N_j}{\sigma_j^2}
=\frac{2N_j-3}{\sigma_j^2}.
$$

\subsubsection{Toy example (continued)}

We consider the same intervention design as before, except that each condition is repeated $40$ times. Let us consider the following design: $\mathcal{J}_{k}=\{1\}$ with $x_1^k=-0.5$ for $k=1\ldots 40$, $\mathcal{J}_{k}=\{2\}$ with $x_2^k=0.5$ for $k=41\ldots 80$, $\mathcal{J}_{k}=\{3\}$ with $x_3^k=0.1$ for $k=81\ldots 120$, $\mathcal{J}_{k}=\{1,2\}$ with $x_1^k=-1.5$ and $x_2^k=2.5$ for $k=121\ldots 160$, and  $\mathcal{J}_{k}=\emptyset$ (no intervention) for $k=161\ldots 200$. We thus have $\mathcal{K}_1=\{41,\ldots,120,161,\ldots,200\}$ with $N_1=120$, $\mathcal{K}_2=\{1,\ldots,40,81,\ldots,120,161,\ldots,200\}$ with $N_2=120$, and $\mathcal{K}_3=\{1,\ldots,80,121,\ldots,200\}$ with $N_3=160$.

With this design, we obtain the following Fisher information matrix for $(\hat w_{1,2}, \hat w_{1,3}, \hat w_{2,3},
\hat \sigma_{1}, \hat \sigma_{2}, \hat \sigma_{3})$:
$$
\mathbf{I}
=\left(
\begin{array}{rrrrrr}
{\bf 27.93981} & 0 & 0 & 0 & 0 & 0  \\
0 &  {\bf 325.4313} & {\bf -291.2836} & 0 & 0 & 0 \\
0 & {\bf -291.2836} &  {\bf 541.3569} & 0 & 0 & 0 \\
0 & 0 & 0 & {\bf 2633.3333} & 0 & 0 \\
0 & 0 & 0 & 0 & {\bf 195.8678} & 0 \\
0 & 0 & 0 & 0 & 0 & {\bf 880.5556} \\
\end{array}
\right)
$$
which is consistent with the inverse of the empirical covariance matrix (sample size $2000$):
$$
\left(
\begin{array}{rrrrrr}
{\bf 27.17689131} &  -1.991505 &  -0.2920015 &  -1.060472  &-0.4260417 & -0.03815717\\
-1.99150498 & {\bf 311.458993} & {\bf -278.4250602} &   2.672117  &-6.1984037  & 4.45498604\\
-0.29200154 & {\bf -278.425060} & {\bf 519.5078012} &  -9.866953   &7.8182677 & -6.64101633\\
-1.06047181 &   2.672117 &  -9.8669528 & {\bf 2708.375488} &-10.5670252 & 13.18525152\\
-0.42604171 &  -6.198404 &   7.8182677 & -10.567025 & {\bf 194.9267470 } & -2.80523059\\
-0.03815717 &   4.454986 &  -6.6410163 &  13.185252  &-2.8052306& {\bf 901.29551952}\\
\end{array}
\right).
$$

\section{Conclusion}

Joint causal network inference from a mixture of observational and intervention transcriptomic data is a very important and challenging research question. In this technical note, we provided an explicit formula for the likelihood function in the context of Gaussian Bayesian networks under any complex intervention design, as well as its analytical maximization. For an unknown graph structure with a known parental node order, it is therefore possible to directly estimate the causal effects. A crucial next step will be to propose an algorithm to obtain the optimal parental order. To this end, we envisage the use of a Mallow's \cite{lu2011learning, doignon2004repeated} proposal distribution in an empirical Bayesian algorithm.

The choice of optimal experimental intervention designs is an important practical question for biologists planning future gene knock-out experiments. Recently, Hauser and B\"uhlmann \cite{Hauser2012} proposed two strategies for the choice of optimal interventions for Gaussian Bayesian networks. The first is a greedy approach using single-vertex interventions that maximize the number of edges that can be oriented after each intervention, and the second yields a minimum set of targets of arbitrary size that guarantee full identifiability. Future research will be needed to determine whether the optimal knock-outs to be performed could alternatively be chosen by evaluating the amount of information potentially contributed by each possible intervention via the Fisher information matrix. We note that the derivation of the Fisher information matrix is not trivial, especially in the case of a mixture of observational and intervention data; in this technical note, we provided  formulae for the calculation of the Fisher information, providing an opportunity for future research concerning optimal experimental intervention designs. 



\end{document}